\documentclass[aps,twocolumn,prd,showpacs,nofootinbib]{revtex4}
\usepackage{amsmath}
\usepackage{graphicx}
\usepackage{dcolumn}
\usepackage{bm}
\usepackage{amssymb}
\usepackage{latexsym}

\def\be{\begin{equation}}
\def\ee{\end{equation}}
\def\ba{\begin{eqnarray}}
\def\ea{\end{eqnarray}}

\begin{document}

\title{On Primordial Density Perturbation and Decaying Speed of Sound}

\author{Yun-Song Piao}

\affiliation{College of Physical Sciences, Graduate School of
Chinese Academy of Sciences, Beijing 100049, China}

\begin{abstract}

The decaying speed of sound can lead to the emergence of the
primordial density perturbation in any expanding phase, even if
the expansion is decelerated. Recently, some proposals have been
given to implement this mechanism, in which it was found that the
primordial spectrum of scalar perturbation can be scale invariant.
In this note, we will give more insights for the details of this
seeding mechanism.

\end{abstract}

\maketitle

In principle, it seems impossible that there is a causal origin of
primordial perturbations in a decelerated expansion phase.
However, if the perturbation is induced by the matter content with
a decaying speed $c_s$ of sound, the conclusion will be altered.
In this case, the nearly scale invariant primordial perturbations
can be surely obtained \cite{Picon}, and \cite{Piao0609}, in which
the parameter $z\sim a$ in the perturbation equation, where $a$ is
the scale factor. In a simple case, the enough efolding number for
the primordial perturbation can be hardly acquired, however, this
embarrassment can be avoided in a slightly complex example
\cite{Piao0609}. Recently, the same mechanism was studied in Ref.
\cite{M}, see also earlier Ref. \cite{PiconL}, in which $z\sim
a/c_s$ in the perturbation equation. In Ref. \cite{M, M1}, the
relation of this seeding mechanism with the varying speed of light
theory \cite{M93, AM} was argued. In Ref. \cite{M2}, there is a
discussion on the connection of the varying speed of sound with
deformed special relativity. It seems that this mechanism can be
interesting, since it not only provides an alternative to the
generation of primordial perturbation, but also may be implemented
in a possible high energy theory.
However, with this in the mind, it will be required, and also
significant to recheck the details of this mechanism, and its
compares with the observations
\footnote{For the thermal fluctuation, e.g. in Refs. \cite{NBV,
MSC}, in order to assure the emergence of primordial perturbation,
the decaying speed of sound has been also applied \cite{Piao0702a,
Piao0702b}.
 }.

We begin with the motion equation of curvature perturbation
induced by the matter content with the speed of sound $c_s$, which
is \be u_k^{\prime\prime} +\left(c_s^2k^2-{z^{\prime\prime}\over
z}\right) u_k = 0 ,\label{uk}\ee see Refs. \cite{ADM, GM}, where
$u_k$ is related to the curvature perturbation $\zeta$ by $u_k
\equiv z\zeta_k$ and the prime denotes the derivative with respect
to the conformal time $\eta$, and $z={a\sqrt{\rho+p}\over c_s h}$,
where $\rho$ is the energy density and $h$ is the Hubble
parameter. For simplicity, we will take $a\sim t^n$ and $c_s\sim
t^p$ for discussion \footnote{Here $c_s$ and the parameter $w$ of
state equation are independent has been implicitly assumed, see
Ref. \cite{Picon, M, M1} for discussions on how having a changed
$c_s$ and an independent relation between $c_s$ and $w$.}, where
both $n$ and $p$ are constants, which are $ a\sim \eta^{n\over
1-n}$ and $c_s\sim \eta^{p\over 1-n}$ after being turned to the
conformal time. In this case, $\sqrt{\rho+p}\sim h$, thus we have
$z\sim a/c_s$.

The general solutions of $\zeta$ given by Eq.(\ref{uk}) are the
combination of $ {\zeta_k }\simeq {c}_1$ and $\zeta_k\simeq
{c}_2\int {d\eta\over z^2}$, where $c_1$ denotes the constant mode
and $c_2$ denotes the mode changed with time. Both are only
$c_sk$-dependent functions. $c_2$ is actually a decreasing mode
for inflation, which is also the reason that we generally do not
consider it in calculating the inflationary perturbation. When the
constant mode $c_1(c_sk)$ is dominated, the spectrum of curvature
perturbation is given by \be {\cal P}^{1/2}_{\zeta}\simeq
{k^{3/2}\over \sqrt{2}\pi}|c_1(\omega)|\simeq f(n){h\over 2\pi
\sqrt{c_s}}(c_sk\eta)^{3/2-v},\label{kuk}\ee where $f(n)\sim {|n-
1|\over \sqrt{2n}}$ is only the function of $n$ and $m_p^2=1$ has
been set. Eq.(\ref{uk}) is a deformed Bessel equation, not like in
usual inflation model, since here $c_s$ is rapidly changed with
the time. Thus in principle $v$ is determined not only by
$z^{\prime\prime}/z$, but also by the evolution of $c_s^2$ before
$k^2$, see Ref. \cite{Piao0609} for some relevant details. Here $v
= {|2p+1-3n|/2|p+1-n|}$ can be obtained by the calculations. The
nearly scale invariance of spectrum requires $v\simeq 3/2$, thus
we have $p\simeq -2$ for all $n$. This means that during the
generation of primordial perturbation the change of $c_s$
corresponds to $c_s\sim (1/t)^2\sim h^2\sim \rho$ \footnote{This
can be implemented in such a model of field theory. We take the
Langrangian of $\varphi$ field as \be {\cal L}=\mu^2_0\sqrt{{\rm
X}}+{\mu_1^6\over \sqrt{\rm X}}-m^2\varphi^2, \ee where ${\rm
X}={(\partial \varphi)^2\over 2}$. For ${\rm X}\gg 1$, the term
${\mu_1^6\over \sqrt{\rm X}}$ is negligible. In this case, this
corresponds to the cuscuton model proposed in Refs.
\cite{Afshordi:2006ad,Afshordi:2007yx}. There is a scale solution
for this Langrangian, which is $h^2\sim \rho\sim \varphi^2\sim
1/t^2$. Thus $\varphi\sim 1/t$ can be obtained, which implies
${\rm X}\sim {\dot \varphi}^2\sim 1/t^4$. Thus we have $\rho\sim
\sqrt{\rm X}$. While the speed of sound is given by \be
c_s^2={{\cal L}_{'{\rm X}}\over {\cal L}_{'{\rm X}} +2{\rm X}{\cal
L}_{'{\rm XX}}}\sim {\rm X}. \label{cs}\ee Thus $\rho\sim c_s$ is
naturally presented. }. This result indicates that in order to
have a nearly scale invariant spectrum $c_s/\rho$ must be
constant, which recently has been pointed out in Ref. \cite{M},
and actually also was implicitly showed in an earlier Ref.
\cite{PiconL} in which one can obtain $c_s^2\sim \rho^2$ from its
Eq.(12) for $\alpha=6$. From Eq.(\ref{kuk}), one can find this
ratio determines the amplitude of primordial perturbation. For any
$n$ not approaching 0 or 1, we need to have ${\cal P}\sim
{\rho/c_s}\sim 10^{-10}$, which is given by the observations
\cite{WMAP5}. In addition, also note that when $p=0$, $v\simeq
3/2$ means that $n$ must approach infinite, thus the usual results
for inflation are recovered. In this case we have $f(n)\simeq
1/\sqrt{2\epsilon}$ since here $\epsilon= 1/n$, thus
Eq.(\ref{kuk}) will exactly equals to that of k-inflation
\cite{ADM, GM}.

The $c_1$ mode is used to calculate the resulting spectrum is
reasonable only when the $c_2$ mode is the decreasing mode. Thus
it is required to examine whether here or in what case the $c_2$
mode is decreased, which was not done in previous references. The
integral$\int {d\eta\over z^2}$ may be straightly calculated,
since $z\sim a/c_s$ is the function of $\eta$. The result
indicates that in order to make the $c_2$ mode decreased with the
time, we need to impose the condition $2p < {3n-1}$. Thus for a
given $n$, if the constant mode is dominated, there must be an
upper bound for $p$. In order to have the scale invariant spectrum
we need $p\simeq -2$, which means that for $0\lesssim n <1$ the
condition $ 2p < {3n-1}$ is always satisfied here. Thus the
validity of Eq.(\ref{kuk}) can be assured.

\begin{figure}[t]
\begin{center}
\includegraphics[width=7cm]{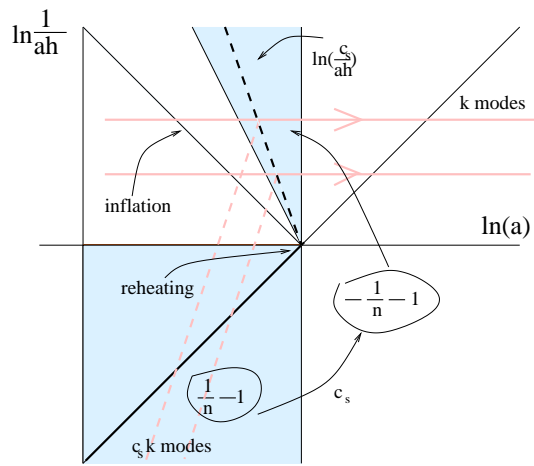}
\caption{ The figure of $\ln{({1\over ah})}$ with respect to
$\ln{a}$, see black solid lines. The red solid lines are the
perturbation modes with wave number $k$. The black dashed line is
that of $\ln{({c_s\over ah})}$ with respect to $\ln{a}$. The blue
region below the $\ln{a}$ axis denotes the decelerated expanding
phase in which $0<n<1$. In principle the primordial perturbation
can not be generated in such a phase. However, when a decaying
$c_s$ is introduced, the effective wave number $c_sk$ of
perturbation mode will not unchange any more, which will decrease
with the time. This under certain condition will make the
evolution of their physical wavelengths able to be faster than
that of $1/h$, see the red dashed lines, thus the corresponding
mode will leave the horizon, see Ref. \cite{Piao0609} for details.
This may be also equally explained as that the perturbation modes
with wave number $k$, see the red solid lines in the left side of
reheating point, leave the sound horizon $c_s/h$. The upper right
region of inflation line is actually the superinflationary region.
}
\end{center}
\end{figure}

The emergence of primordial perturbation can be explained as
follows. In the regime $c_sk\eta \gg 1 $, we have $a/k\ll c_s/h$.
Thus though $a/k\gg 1/h$, i.e. the physical wave length of
perturbation mode is larger than the horizon, it is actually
smaller than the sound horizon $c_s/h$ since the speed $c_s$ of
sound is large. Thus a causal relation can be established on
superhorizon scale. When $c_s$ is decreased the corresponding mode
will leave the sound horizon and can be able to be responsible for
the seed in observable universe. In this case the horizon $1/h$ in
the conventional consideration should be replaced by the sound
horizon $c_s/h$. The scale invariant spectrum requires $c_s\sim
h^2$. Thus this replacement can be equally written as
$1/h\rightarrow h$, since $c_s/h\sim h$. We generally have $a \sim
({h\over a})^{-{n\over 1+n}}\sim ({c_s\over ah})^{-{n\over 1+n}}$.
Thus Eq.(3) in Ref. \cite{Piao0609} is replaced as \be
\ln{({c_s\over ah})}=-({1\over n}+1)\ln{a}, \label{ah3}\ee which
is plotted in Fig.1 as the black dashed lines. Thus in principle
the rapid decaying of $c_s$ is included for a region of
decelerated expansion, see the blue region below the $\ln{a}$
axis, corresponds to map this region to a dual superinflation
region, see the same blue one beyond the $\ln{a}$ axis, by a dual
replacement $1/h\rightarrow c_s/h\sim h$. The results in Eq.(3) in
Ref. \cite{Piao0609} and Eq.(\ref{ah3}) are just a reflection of
this duality, which has been plotted in Fig.1. This duality is not
illustrated in Ref \cite{Piao0609}. For example, for an expansion
dominated by radiation, we have $n=0.5$ and thus $\ln{({1/
ah})}=\ln{a}$, see the black thick solid line in Fig.1. When
$c_s\sim h^2$ is introduced, with Eq.(\ref{ah3}), we have
$\ln{({c_s/ ah})}=-3\ln{a}$, see the black dashed line in Fig.1.

The efolding number can be defined as the ratio of the physical
wavelength corresponding to the present observable scale to that
at the end time of the generating phase of perturbations, which in
general is actually not the efolding number of scale factor, but
is that of primordial perturbation.
The efolding number is \cite{Piao0609} \be {\cal N} =
\ln({c_s\over c_{s(e)}}\cdot {k_e\over k})= (n+1)\ln({h\over
h_e}), \label{caln2}\ee where the subscript `$e$' denotes the end
time of the generating phase of perturbations, and $h\sim
1/a^{{1\over n}}$ has been used. Here $k=ah$ and $c_s\sim h^2$,
thus during this seeding mechanism ${\cal N}$ is actually equal to
the change of $\ln{(h/a)}$, which is consistent with
Eq.(\ref{ah3}) and also the black dashed line in Fig.1. The
prefactor of Eq.(\ref{caln2}) is distinguished from that of
Eq.(12) in Ref. \cite{Piao0609}. The reason of difference is that
there we straightly introduce a scalar field with separating time
derivative and space derivative, which leads to that $z\sim a$
defined in Eq.(\ref{uk}) is independent of $c_s$, while here
$z\sim a/c_s$. In Ref. \cite{Piao0609}, it was showed that in a
simple case the corresponding model can hardly have enough
efolding number. We will see, however, that this difficulty dose
not exist here. The key point is here to obtain the scale
invariance of spectrum $c_s$ has to decrease more rapidly than
that in Ref. \cite{Piao0609}, which thus will lead to larger
efolding number by Eq.(\ref{caln2}).

\begin{figure}[t]
\begin{center}
\includegraphics[width=7cm]{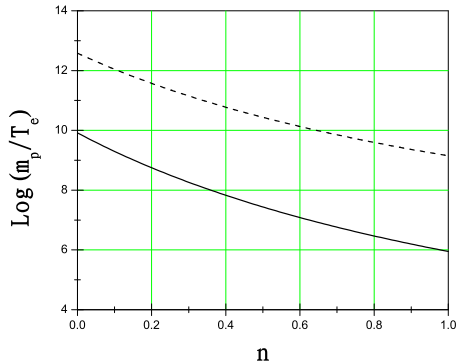}
\caption{ The figure of the $\log ({m_p\over T_e}) $ with respect
to $n$ showing how enough efolding number is obtained. The solid
line is for the case that the energy scale $\sim m_p$ at the
beginning time of phase generating the primordial perturbations
and the dashed line is that of the energy scale $\sim m_p/10^4$.
The region above the corresponding line is that with enough
efolding number. }
\end{center}
\end{figure}

In Eq.(\ref{caln2}), the resulting $\cal N$ depends on the ratio
$h$ to $h_e$, which must be large enough to match the requirement
of observable cosmology. $\cal N$ required is generally determined
by the evolution of standard cosmology after reheating. Here the
``reheating" means that the matter content generating the
primordial perturbation, e.g. that in footnote 3, decays into
radiation. In principle the lower $T_e$ is, where $T_e$ is the
reheating scale, the smaller the efolding number required is. For
an idealistic case, in which after the generating phase of
perturbations ends the universe will rapidly be linked to an usual
evolution of standard cosmology, $ {\cal N} \simeq 68.5+
0.5\ln(h_e/m_p)$ can be obtained \cite{LL}, which actually
approximately equals to ${\cal N}\simeq 36+\ln{({T_e\over {\rm
Tev}})}$ given by Ref. \cite{WMAP5}.
We can substituting it into (\ref{caln2}) to cancel ${\cal N}$,
and obtain a relation between $h_e$ and $n$. We plot Fig.2 in
which for various $n$ in the region $0\lesssim n<1$, the $\log
({m_p\over T_e}) $ required for enough efolding number is given,
where $m_p$ is the Planck scale.
We can see from Fig.2 that when taking the initial energy scale as
the Planck scale or grand united scale, enough efolding number can
be easily obtained, since in principle $T_e$ may low to the
nucleosynthesis scale in which $\log({m_p\over T_e})\simeq 22$
while in above two cases $\log({m_p\over T_e})\lesssim 13$. For
example, for $n=0.5$, we have $T_e\sim 10^{12}$Gev for $m_p$ and
$T_e\sim 10$Tev for grand united scale, respectively.

In conclusion, it is rechecked that the primordial density
perturbation can be generated in a decelerated expanding phase
with decaying $c_s$. We give more insights for this seeding
mechanism, and show that, for this mechanism, the enough efolding
number of primordial perturbation can be obtained. However, it
should be mentioned that this mechanism only serves the generation
of primordial perturbation, it can not solve all problems of
standard cosmology, as has been explained in inflation scenario.
Thus unless there are some other mechanisms to give the solutions
of above problems, it seems inevitable that we still needs a
following period of inflation. However, in the latter case, the
seeding of perturbation and the stage of inflation may be actually
decoupled, which might help to relax the bounds for inflation
model itself leaded by the observations.
In some sense, this work may be interesting for coming endeavor of
embedding such a seeding mechanism into a possible fundamental
theory.

\textbf{Acknowledgments} This work is supported in part by NSFC
under Grant No: 10775180, in part by the Scientific Research Fund
of GUCAS(NO.055101BM03), in part by CAS under Grant No:
KJCX3-SYW-N2.

\end{document}